# Single organic molecules for photonic quantum technologies

C. Toninelli[1], I. Gerhardt[2], A.S. Clark[3], A. Reserbat-Plantey[4], S. Götzinger[5,6], Z. Ristanovic[7], M. Colautti[1], P. Lombardi[1], K.D. Major[3], I. Deperasińska[8], W.H. Pernice[9], F.H.L. Koppens[4], B. Kozankiewicz[8], A. Gourdon[10], V. Sandoghdar[5,6], and M. Orrit[7]

1 CNR-INO and LENS, Via Nello Carrara 1, 50019 Sesto Fiorentino (FI), Italy
2 Institute for Quantum Science and Technology (IQST) and 3rd Institute of Physics, D-70569 Stuttgart, Germany
3 Centre for Cold Matter, Blackett Laboratory, Imperial College London, Prince Consort Road, SW7 2AZ, London, United Kingdom
4 ICFO − Institut de Ciencies Fotoniques, The Barcelona Institute of Science and Technology, 08860 Castelldefels (Barcelona), Spain
5 Max Planck Institute for the Science of Light, 91058 Erlangen, Germany;
6 Friedrich-Alexander University of Erlangen-Nürnberg, Germany;
7 Huygens-Kamerlingh Onnes Laboratory, LION, Postbus 9504, 2300 RA Leiden, The Netherlands
8 Institute of Physics, Polish Academy of Sciences, Al. Lotnikow 32/46, 02-668 Warsaw, Poland
9 Physikalisches Institut, Westfälische Wilhelms, Universität Münster, Heisenbergstrasse 11, 48149 Münster, Germany
10 CEMES-CNRS 29 Rue J. Marvig, 31055 Toulouse, France

**Isolating single molecules in the solid state has allowed fundamental experiments in basic and applied sciences. When cooled down to liquid helium temperature, certain molecules show transition lines, that are tens of megahertz wide, limited only by the excited state lifetime. The extreme flexibility in the synthesis of organic materials provides, at low costs, a wide palette of emission wavelengths and supporting matrices for such single chromophores. In the last decades, the controlled coupling to photonic structures has led to an optimized interaction efficiency with light. Molecules can hence be operated as single photon sources and as non-linear elements with competitive performance in terms of coherence, scalability and compatibility with diverse integrated platforms. Moreover, they can be used as transducers for the optical read-out of fields and material properties, with the promise of single-quanta resolution in the sensing of charges and motion. We show that quantum emitters based on single molecules hold promise to play a key role in the development of quantum science and technologies.**

Modern societies have an ever-growing need for efficient computation techniques and for fast and secure communication, to distribute a huge amount of data around the globe. By harnessing quantum effects present at the nanoscale, new quantum technologies can be employed to meet these needs, including quantum cryptography and fully-fledged quantum information processing. On the other hand, the extreme sensitivity of quantum systems to their local environment can be exploited to also create new sensing devices, which provide unprecedented precision, accuracy and resolution and can be deployed within large quantum networks.

Key applications require the generation and manipulation of quantum states of light, such as photonic quantum simulation [1, 2], linear optical quantum computing [3], device-independent or long-distance quantum key distribution protocols [4], sub-shot-noise imaging [5] and quantum metrology [6, 7]. In this context, single impurities in solid-state systems can act as bright, on-demand single-photon sources (SPSs), which are a crucial resource in these photonic quantum technologies. Quantum emitters may also perform as non-linear elements at the few-photon level [8] and as nanoscale sensors, allowing the optical read-out of local properties of materials and fields. In this context, single molecules in the solid-state offer competitive and reliable properties, with several key advantages. First, they are very small and have well-defined transition dipole moments so that they can be used as nanoscopic sensors for a number of scalar and vector quantities such as pressure, strain, temperature, electric and magnetic fields, as well as optical fields. Second, organic molecules can be designed and synthesized for different parts of the visible spectrum and integrated in a range of organic matrices, a feature that is a severe limiting factor for color centers in diamond or lithographically produced semiconductor quantum dots. Third, the combination of small size and ease of fabrication makes organic molecules ideal for applications where high densities and scalability are desirable. Fourth, organic dye molecules can have strong zero-phonon lines, which reach their Fourier-limited natural linewidth at liquid helium temperature, thus providing very bright and stable sources of photons with a high degree of coherence.

Although many of these features have been known since the early nineties, thanks to the development of specific, efficient light-molecule interfaces and hybrid molecular devices in the last ten years, molecular quantum emitters are making the jump from proof-of-principle to the era of complex quantum optics experiments and multi-photon devices.

Here we review the recent advances of single-molecule studies for quantum technologies, with a special focus on the coupling to nanophotonic structures for the enhancement and control of their interaction with quantum light. We discuss how this can extend the quantum optical toolbox for molecules, enabling flexible and efficient quantum photonic devices and new diverse applications.

In Section 1 we discuss some fundamentals of molecules and their photophysics. For a general overview in the field of single-molecule physics and spectroscopy, we refer the reader to review papers such as Refs. [9,10,11]. Some key experiments of the past decade are then summarized, with single molecules acting as well-isolated single quantum systems: we review the advances in the generation of non-classical states of light in Section 2, while nonlinear light-matter interactions at the single-molecule level are reported in Section 3. Section 4 highlights recent results in the integration of molecules into photonic structures for efficient light extraction and the development of integrated architectures for experiments with increased complexity. In Section 5 we present progress in quantum sensing with molecules, and finally discuss future applications and an outlook for molecular quantum technologies.

# 1. Basics of single-molecule photophysics

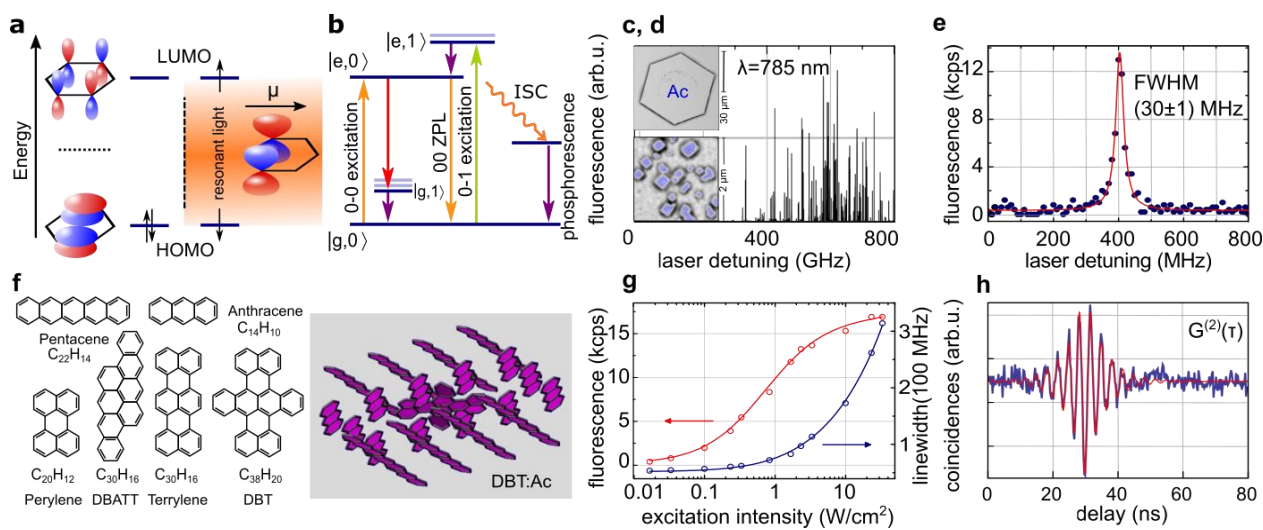

**Figure 1: Single-molecule photophysics. a** Artistic view of aromatic molecular orbitals coupled by a resonant light field, resulting in an induced dipole moment μ. **b** Simplified Jabłoński diagram showing the molecular energy levels which are relevant in the fluorescence process. The eigenstate notation describes the excitation of both the electronic and the vibrational degrees of freedom (just one mechanical mode is depicted for clarity). Purple lines correspond to non-radiative transitions, ISC and ZPL stand for intersystem crossing and zero-phonon line, respectively. **c** Scanning electron microscope images of anthracene crystals obtained via sublimation and reprecipitation [20]. **d** Fluorescence excitation spectrum of a Dibenzoterrylene:Anthracene crystal, showing the expected inhomogenous broadening, spanning a few hundred GHz around a central wavelength of 785 nm. **e** Fluorescence excitation spectrum of a single DBT molecule in naphthalene at 1.4 K revealing the Lorentzian profile of the ZPL, with a Fourier-limited linewidth of 30-MHz FWHM. **f** Chemical structures of the most relevant PAH molecules in quantum technologies. Right: results of DFT calculations for the Ac crystal structure with a DBT molecule replacing three host molecules, using the Gaussian 09 package. **g** Power dependences of the fluorescence emission and of the ZPL linewidth from a single DBT molecule [20]. The coherence of the electronic ZPL transition allows for the observation of several Rabi cycles in panel **h**, displaying the autocorrelation function from a start-stop measurement, upon resonant pumping. Panels **c**, **d** and **g** are adapted with permission from Ref. [20], copyright (2018) American Chemical Society.

Many organic molecules have electronic transitions in the spectral range from 200 to 800 nm. This transition is associated with the light-induced dipole moment, which results from the coupling between the lowest unoccupied molecular orbital (LUMO) and the highest occupied molecular orbital (HOMO) (Fig. 1a), and has a strength of a few Debye. Fluorescence is the emission of light due to electronic transitions between singlet states following an absorption event, with typical excited state lifetimes of a few nanoseconds. A vibrational manifold accompanies the electronic levels (Fig. 1b) and due to electron-phonon coupling part of the emission is red-shifted compared to the excitation light, meaning they can be separated with standard optical filters. After the first demonstration of single-molecule fluorescence detection at cryogenic temperature [12], this almost background-free technique has enabled single-molecule experiments at room temperature in a wide spread of disciplines such as biology [13] and material science [14].

Differences in the vibrational potential of the electronic ground and excited states mean the motion of the nuclei is coupled to the electronic transitions. The larger the similarity between the nuclear positions in the ground and excited state, the stronger the transition probability for the so-called 0-0 zero-phonon line (ZPL), that does not involve the excitation of molecular vibrations. In particular, for well-studied rigid dye molecules such as terrylene (Tr), dibenzanthanthrene (DBATT), and dibenzoterrylene (DBT), the energy stabilization upon nuclear relaxation in the excited state is as small as a few tens of $cm^{-1}$ (several meV) and the emission into the 0-0 ZPL at low temperature occurs with a probability exceeding 50%. Considering the vibronic coupling in the linear approximation, the Franck-Condon factors account for the relative intensity of the different vibro-electronic transitions, associated to a given harmonic vibrational mode. Room-temperature single-molecule spectroscopy with ultrafast laser pulses can unveil the coherence between these levels [15,16].

Intersystem crossing (ISC in Fig. 1b) is a process that competes with fluorescence. This spin-forbidden transition is allowed by spin-orbit coupling, a weak interaction as long as no heavy atoms are involved. The molecule is then trapped in a long-lived triplet state until emission occurs, which in this case is called phosphorescence. ISC is important not only in solar cells and display technology, but also for the electrical excitation of single molecules [17,18]. Triplet states can also be used for interfacing molecules with magnetic fields and microwaves, which we discuss in Section 5.

To maximize the fluorescence quantum yield and emission brightness, guest-host systems must be carefully selected to have a low ISC probability, which often requires that the excited states of the host matrix, both singlet and triplet, are located energetically above the guest's excited singlet state [19].

With these requirements in mind, single molecules can be conveniently embedded as substitutional impurities in crystalline matrices with diverse fabrication processes. These result in crystals with different dimensions and aspect ratios depending on the method and on the host system (see Supplementary Information). Figure 1c, for example, shows two scanning electron microscope images corresponding to anthracene crystals obtained by sublimation and reprecipitation [20]. In these solid materials, the electron-phonon coupling also involves the surrounding host system, which gives rise to a broad phonon wing in the emission spectrum, red-shifted with respect to the common mode of the 0-0 ZPL [21]. The Debye-Waller factor describes the extent of this interaction and is defined as the ratio of the intensity emitted into the 0-0 ZPL to the total emitted intensity. This factor decays exponentially with temperature [10], resulting in a negligible emission in the 0-0 ZPL for molecules at ambient conditions. Such coupling with phonons, depending on the nuclear potential landscape, also induces the so-called Stokes shift between the energy of the maximum in the absorption and emission spectrum.

Moreover, second-order phonon interactions bring about dephasing of the induced dipole and hence cause a temperature-dependent broadening of the 0-0 transition [21, 22]. For certain guest-host systems and very rigid molecules under cryogenic conditions (below 2K), the ZPL narrows down to its natural linewidth, typically of the order of a few tens of MHz. This sets them apart from molecules such as rhodamines and coumarins which are very common fluorescent labels but display broad lines under cryogenic conditions. Although the individual guest molecules are nominally all identical, their transition frequencies are also influenced by local charges and strain in their environment, which can be slightly different from point to point inside the matrix. This results in a residual inhomogenous broadening that can though be as small as few GHz in high-quality sublimation-grown crystals (see Fig.1d), as for instance reported for pentacene in *p*-terphenyl [23]. More typical values of the inhomogeneous width are tens of GHz, and can reach thousands of GHz in non-equilibrium crystallizations [24] or amorphous hosts such as polymers [25].

In the domain of quantum optics, where a high degree of coherence and thus lifetime-limited emission is crucial, planar and rigid polycyclic aromatic hydrocarbons (PAHs) are of particular interest because their linewidths tend to be dramatically reduced at temperatures below 2 K [10] (Fig.1e). The most commonly used guest PAH molecules include pentacene, terrylene, perylene, dibenzanthanthrene (DBATT) and dibenzoterrylene (DBT), whose chemical structures are shown in Fig. 1(f). Host molecules are often chemically saturated (i.e., they present only single C-C bonds), but can also be light PAHs such as substituted benzene, terphenyl, naphthalene, anthracene (Ac), and some of their derivatives. The molecular insertion of the guest in the host lattice, corresponding to the minimum free energy, can also be calculated by molecular mechanics. An example is sketched in Fig. 1f for DBT in Ac. Once inserted into a suitable host, these fluorescent guest molecules exhibit a small Stokes shift and a fluorescence quantum yield (QY) close to unity [26]. Moreover, for PAHs well embedded in a crystal at low temperature, photochemistry is usually absent, so that they are expected to be indefinitely photostable under normal illumination conditions [27].

Under resonant excitation (0-0 excitation, red arrow in Fig. 1b) of the electronic transition at cryogenic temperature, red-shifted photons are usually detected after spectral rejection of the pump light. The decay of the excited state into the vibrational manifold of the electronic ground state also implies that consecutive emitted photons are not coherent. However, the coherence of the molecules' internal states in the laser field before the emission is accessible (Fig. 1h) [28]. When the frequency of a narrow-linewidth laser is tuned across the transition, the detected fluorescence counts display a typical Lorentzian line shape, associated with the exponential spontaneous decay of the excited state (Fig. 1e). As the illumination intensity of the laser is increased, the excitation line broadens and the peak count rate saturates, (Fig.1g). The single-photon nature of the emitted light was further explored in Ref. [29], where resonance fluorescence was observed together with the well-known Mollow triplet, due to Rabi oscillations in the strong laser field. Pulsed excitation of a molecule has also allowed coherent state preparation, and up to $11\pi$ Rabi cycles were observed [30].

When lifetime-limited photons are required, it is possible to use a cross-polarization configuration to suppress the excitation light [29], or excite the molecule into a higher vibrational level of the electronic excited state. This so-called 0-1 excitation allows for the generation of spectrally narrow-band photons, which have the linewidth of the 0-0 transition [31, 32, 33, 34]. This is particularly relevant for the generation of coherent, indistinguishable photons, discussed in the next section.

# 2. Single-molecule-based single-photon sources

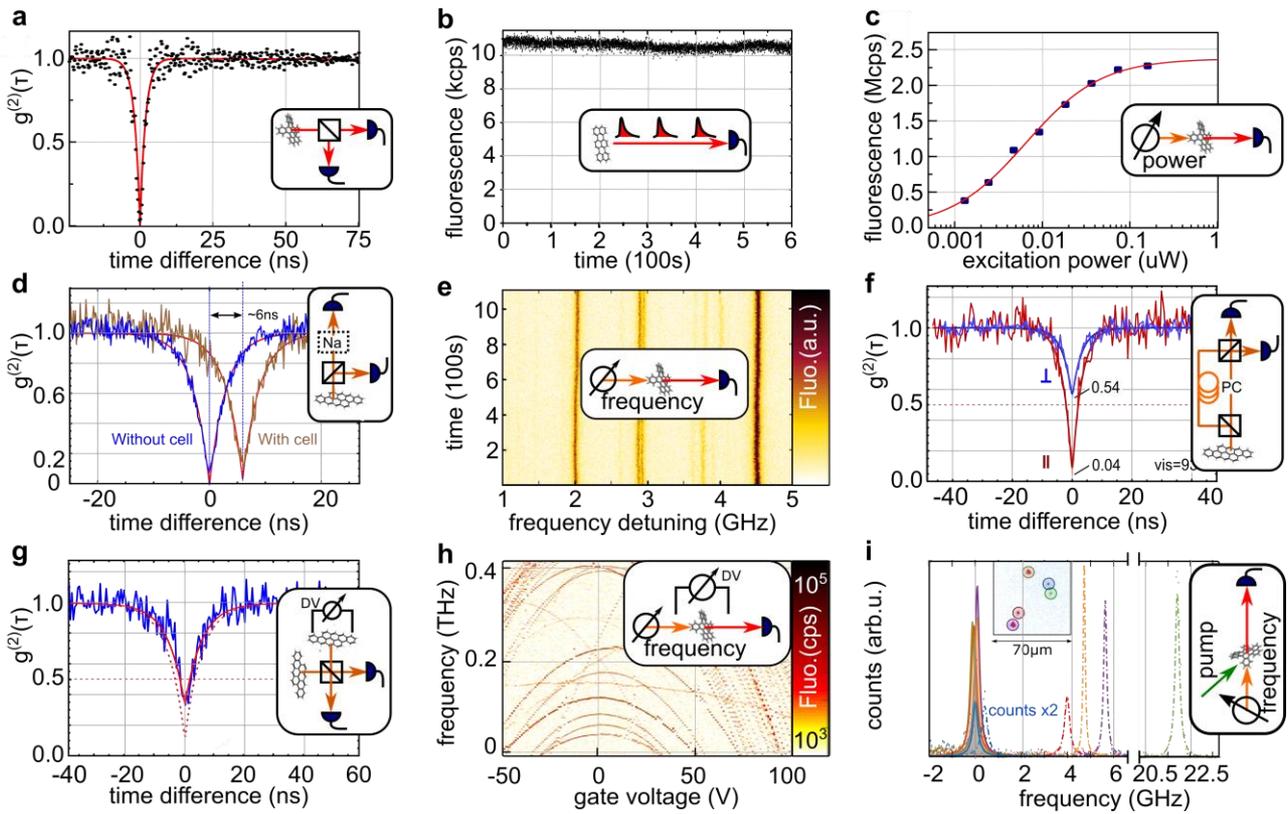

**Figure 2**: **Single-molecule-based single-photon sources**, **a** Example of a second-order autocorrelation function, measured in a Hanbury Brown and Twiss configuration for a single DBT molecule in Ac, upon CW 0-1 pumping [49]. **b** Typical time trace obtained for single-molecule emission at room temperature for terrylene in *p*-terphenyl. **c** Saturation curve showing the detected, back-ground corrected counts from a single molecule at low temperature, as reported in Ref [49]. **d** First reported interaction between single photons from molecules and atomic systems [52]. **e** Excitation spectra for DBT molecules in 2,3-dibromonaphthalene shown in color scale as a function of time, illustrating negligible spectral diffusion [60]. **f** Two-photon interference obtained in a Hong-Ou-Mandel configuration for the photon stream from a single DBATT molecule as reported in Ref [42]. **g** Two-photon interference in the photon streams obtained collecting the ZPLs of two remote single molecules under CW 0-1 pumping [31]. **h** Excitation spectra are reported in color scale as in e), showing the characteristic second-order Stark shift of DBT molecules, induced by the applied gate voltage. The sample is described in Ref [61]. **i** Excitation spectra of five different DBT molecules whose relative position is given in the confocal fluorescence map in the inset [63]. The ZPL frequency of four of them is brought in resonance (within 2 linewidths) with the fifth one (solid-blue filled line), by means of a superimposed pump beam. Dashed and solid lines refer to the spectra before and after tuning, respectively. The insets to all panels represent the measurement concept and the employed molecular source. Orange and red solid lines correspond to light at the ZPL frequency and in the red-shifted channels, respectively [63]. Panels **a c** adapted with permission from Ref [49] Copyright 2020, Wiley-VCH; **d** from ref [52], **e** from [60], **f** from Ref [42], **g** from Ref [31], **h** from Ref [61] copyright (2019) American Chemical Society, **i** from Ref [63]

In this section, we focus our discussion in the use of molecules as single-photon sources. In the condensed phase, single molecules were the first systems to show single-photon statistics. This was observed for pentacene in *p*-terphenyl at liquid helium temperatures as early as 1992 [28]. Later these observations were extended to different molecules in liquids and on surfaces at ambient conditions [35, 36, 37]. The state-of-the-art performance of molecule-based single photon sources is summarized in Table 1, which shows values competitive with the best available solid-state sources, (see e.g. Refs. [38, 39, 40]).

The property of emitting one photon at a time, ideally on demand, is relevant for many quantum technology applications. Any background alters the photon statistics and spoils the performance of such single-photon-based devices. The single-photon purity is the ratio of the probability of generating a single photon, $p_1$, over the joint probability of there being any photon event $\sum_n p_n$, where $p_n$ is the probability of detecting $n$ photons [41]. This is commonly estimated by measuring the auto-correlation of the emitted light, $g^{(2)}(t)$, as $1 - g^{(2)}(0)/2$. Indeed, for low mean photon number, the following relationship holds $g^{(2)}(0) = 2p_2/p_1^2$. For an ideal single-photon emitter, $g^{(2)}(0) = 0$, a feature referred to as "anti-bunching", and the purity is 1. Molecules can achieve $g^{(2)}(0) = 0.00(3)$ when finite detector timing resolution is taken into account [42], and hence are competitive with other solid-state systems (Fig. 2a).

A further important and related figure of merit of a single-photon source is its efficiency, often termed *brightness*. The efficiency with which a single quantum emitter can generate photons in a known optical mode depends on a number of factors, including the efficiency of excitation (absorption cross section), the quantum yield of fluorescence, and the overall collection efficiency. Each of these terms has to be optimized. By employing 0-1 excitation for instance, the population of the molecule can be fully inverted at low temperature, corresponding to unity excitation efficiency [43].

Several common PAH molecules display a near-unity quantum yield (see Ref. [26]) and show negligible ISC meaning they exhibit minimal blinking. In the case of DBT in anthracene for instance, the ISC probability at low temperature is as low as $10^{-7}$ [44]. This source, emitting in the near infrared at around 785 nm, is stable both at room [45, 46] and at cryogenic temperatures [47] with an off-time below a per cent. A time-trace with microsecond binning time, recorded at ambient conditions from a single terrylene molecule in a *p*-terphenyl crystal is shown as an example in Fig. 2b. Such a source, combined with smart photonic engineering, yields an extremely regular stream of single photons [48], leading to intensity squeezing of up to 2.2dB at room temperature.

The collection efficiency is typically the most limiting factor in the overall loss budget for molecule-based single-photon sources (see Table 1). Simple photonic structures (see Section 4) can improve collection efficiencies from a few per cent with low numerical optics, up to 20% into a single spatial mode even in cryogenic experiments [49]. Near-unity collection efficiency in ambient conditions has been demonstrated for the combination of a planar antenna with high numerical aperture optics [50].

The product of the efficiency and the rate at which one can repeatedly excite the system gives a number of *photon counts per second*. For single molecules, this is comparable to the best single-photon sources, with just under 50 million photons per second being detected from a single molecule at room temperature [50]. Detected count rates at cryogenic temperatures exceeding a few million counts per second under continuous-wave 0-0 excitation have been demonstrated [49] (Fig. 2c), reaching the order of ten million counts per second for a single molecule coupled to a plasmonic nanoantenna [51].

Considering a ZPL branching ratio of up to 50% from the product of the Debye-Waller and Franck-Condon factors, starting from an excited-state lifetime of about 5 ns, a 100% collection efficiency would correspond to a count rate of 100 MHz at the first lens. Such values for the collection efficiency are within reach in certain photonic structures, as discussed in Section 4.

Molecule-based single-photon sources can also be operated under pulsed excitation [37, 47, 46], and recent efforts to optimize collection efficiency by coupling molecules to photonic structures can be readily extended to the pulsed domain.

Under cryogenic conditions, both 0-0 excitation and 0-1 pumping schemes can provide lifetime-limited emission [34, 42]. The ultra-narrow linewidth of the 0-0 ZPLs in PAHs, typically only a few tens of MHz wide (Fig. 1e), offer a unique advantage for the efficient coupling of single photons to other emitters or even to cold atomic systems. In Ref. [52], for example, the authors demonstrate the interaction of photons originating from a single dibenzanthanthrene molecule with sodium atoms in the gas phase (Fig. 2d).

Moreover, for some solid-state systems the emission of narrow-linewidth photons is hindered by frequency fluctuations on short time scales, often called spectral diffusion. In the case of very rigid matrices and well-behaved PAH emitters, negligibly small spectral diffusion can be achieved even with simple sample preparation procedures. For example, a dilution of DBT in melted, polycrystalline naphthalene, leads to highly stable ZPLs with nearly no spectral diffusion. A similar effect is achieved in host matrices with halogen heteroatoms. Figure 2e shows real-time excitation spectra of DBT in 2,3-dibromonaphthalene, illustrating high spectral stability over about 20 min. The sensitivity of the ZPL frequency to temperature and pressure, leading to possible fluctuations, are extensively discussed in Section 5.

An important figure of merit for SPSs is the ability to generate photons that undergo two-photon interference (TPI) when impinging on separate input ports of a 50:50 beam splitter. This so-called Hong-Ou-Mandel interference effect has been used to enable precision relative timing measurements [53] and is the central technique to enable Bell-state analysis for quantum teleportation [54, 55], as well as linear optical quantum computing (LOQC) schemes [3]. Furthermore, the remote interference of two quantum emitters can be used to prepare them in a joint entangled state. After the first experiments on TPI [56, 32], by using an atomic filter on single DBATT emission, a visibility of $0.94 \pm 0.03$ was achieved [42] (Fig. 2f). TPI has also been observed for photons from two remote molecules [31]. This impressive result, which pioneered such measurements in solid-state quantum emitters, was made possible by the small inhomogeneous broadening that can be achieved in high-quality molecular crystals (tens of GHz) and the local tunability of molecules through the Stark effect [57]. Finally, two independent photons were entangled using a beam-splitter, proven by a Bell-violation in the raw data of $2.24 \pm 0.12$ [58].

Depending on the internal structure, the Stark effect is generally found to be linear for molecules without inversion symmetry, or quadratic for inversion-symmetric molecules [59]. However, local imperfections in the host matrix can lift the degeneracy, such that a linear Stark shift is observed even in nominally centro-symmetric impurities. In this way a sensitivity of 1.5 GHz/(kVcm$^{-1}$) was recently achieved [60]. Naturally, electric charges and fields can be probed by this effect – even on nanoscopic length scales, as discussed below in Section 5. In Fig. 2h we report in color scale the excitation spectrum of several DBT molecules in anthracene as a function of the applied gate voltage, reaching a total shift of more than 200 GHz [61]. Here DBT:Ac nanocrystals were integrated in a polymer matrix which formed a substrate for the transfer of a 2D-material electrode.

A new method is emerging to optically tune single molecule ZPLs without the introduction of any nano-fabrication step. This is based on an additional pump beam which induces locally a persistent electric field, as a function of the applied dose. This approach has allowed matching the resonance of up to five emitters within twice their linewidth (Fig. 2i), all within an area of about 50 microns. With a resolution mostly determined by the laser spot size, it appears particularly suitable for molecular emitters integrated on chips [62].

In conclusion, the molecular systems presented here are state-of-the art single photon sources and excel in terms of optical coherence time, intensity and frequency stability of the photon stream and also with respect to the possibility of interfering photons from distinct molecules. These features, combined with the ease of fabrication, tuning and integration (see also Sections 4 and 5) will be relevant to speed-up multi-photon devices and hence enable the implementation of small scale linear-optical quantum computers and long distance QKD.

| Figure of merit | Definition and measurement strategies | State-of-the-art for molecules | | References |
|---|---|---|---|---|
| Collection Efficiency & max count rate | Probability that a generated photon is collected per trigger event & maximum count rate at the detector | RT | free space (N.A.1.65): >96%, waveguide: ~20% | [50, 48] [63] |
| | | LT | free space (N.A.0.67): ~40%, ~ 4 Mphotons/s waveguide: ~5% | [49] [64] |
| Single-Photon Purity | $p_1 / \sum_n p_n \to 1 - g^{(2)}(0)/2$ Second-order correlation function | RT | $g^{(2)}(0) = 0.03^{+0.01}_{-0.01}$ | [20] |
| | | LT | $g^{(2)}(0) = 0.00^{+0.00}_{-0.03}$ | [42] [49] |
| Linewidth | Resonant laser scans or tunable filtering | RT | ~30 THz | [45] |
| | | LT | ~10-100 MHz | [65] |
| Indistinguishability | Two-photon interference visibility | 1 mol | ~ 95% | [42] |
| | | 2 mol | ~ 75% | [31] |
| Tunability | Voltage-dependent excitation spectra | (LT): 300 MHz/(kV cm$^{-1}$) linear, 0.15 MHz/(kV cm$^{-1}$)$^2$ quadratic, 400 GHz tuning range | | [61] |
| | | (LT): 1.5 GHz/(kVcm$^{-1}$), linear response | | [60] |

**Table 1 | Figures of merit characterizing a single-photon source (SPS)** including efficiency, single-photon purity, linewidth of the associated radiative transition, photon indistinguishability, and frequency tunability. We here summarize these concepts and the state-of-the-art values for molecules. The count rate at the detector is normalized by the detector quantum efficiency only; RT and LT stand for room temperature and low temperature respectively. The maximum count rate detected is not mentioned for the waveguide-integrated molecules because it depends on the out-coupler efficiency and is not relevant here.

# 3. Single-molecule optical nonlinearities

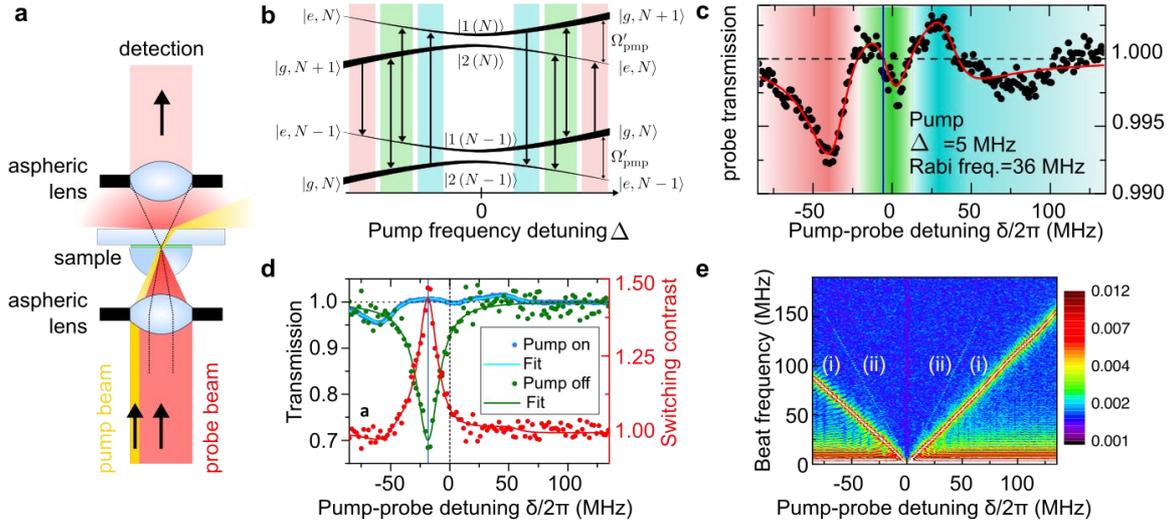

**Figure 3**: **A single molecule acting as a nonlinear medium; a** Sketch of a setup for transmission measurements on a single molecule subjected to superimposed probe and pump beam, inducing a nonlinear behavior. **b** Simplified level scheme of a single molecule under the influence of a strong pump beam. Shown are the new eigenstates of the combined molecular resonance and the pump beam, as a function of the pump detuning. The thickness of the energy levels indicates the difference in population. The colors indicate the transitions at probe frequency $\omega_{pump} - \Omega'_{pump}$ (red), $\omega_{pump}$ (green) and $\omega_{pump} + \Omega'_{pump}$ (blue). **c** Transmitted power of a weak probe beam. The underlying colors indicate different nonlinear processes. **d** Switching the transmission of the probe beam with a weak pump beam. The red curve is obtained as the ratio between the probe transmission with pump on and off. **e** Fourier transform of the time-dependent measurement recorded in transmission. The colour shading indicates the magnitude of the Fourier transform of the transmission signal on a logarithmic scale. The straight line marked with (ii) is the result of a four-wave mixing process. The generated light interferes with the probe beam creating a beat frequency. Panel **c**, **d** and **e** are adapted with permission from ref. [71], add publisher. Panel **b** is adapted with permission from [69]

In recent years the desire to control photons with photons has flourished, spurred on by the ever-increasing demand for low-power classical transistors and novel potential applications of optical quantum information processing. Photon-photon interactions can be effectively mediated by matter. If the interaction between light and matter is strong enough, a single photon can saturate a single emitter such that the reaction of the system to one or two photons is very different. The large extinction cross-section of a single molecule at cryogenic temperatures, together with the possibility to tightly focus a laser beam using high refractive index solid immersion lenses, offers the possibility to observe saturation at the few-photon level. About a decade ago it was demonstrated that a single molecule can act as an optical transistor [66]. A single molecule is sufficient to reflect about 20% of a weak probe laser beam when tightly focused onto the molecule [67]. A setup for transmission measurements is depicted in Fig.3a. The situation changes when a second pump laser excites the molecule into a vibrationally excited state of the first electronic excited state. The intensity of the transmitted probe light is now dependent on the power of the pump laser. For larger excitation powers, stimulated emission even amplifies the transmitted laser beam. Many pump photons are required in this configuration to control the transmitted laser beam, due to the low absorption cross section of the vibrationally excited state. Ultimately, the goal is to control one photon using a second photon, via the controlled interaction with a single molecule. The key to this goal is the strong nonlinearity, which is demonstrated by the saturation behavior described above, combined with nanophotonic structures, which we describe in Section 4.

The use of two beams with different frequencies gives access to other nonlinear effects beyond simple saturation. Indeed, the interaction of ensembles of atoms with two laser beams (a pump and a probe) was studied extensively in the 1970s. Effects like the AC-Stark shift, stimulated Rayleigh scattering and the hyper-Raman effect were first observed with ensembles of atoms [68] and later by monitoring the fluorescence, even with a single molecule [68]. All these effects can be understood in the dressed-atom picture, where the pump laser induces a split of the levels by the effective Rabi-frequency. The populations of the new dressed eigenstates depend on the detuning of the pump laser with respect to the transition of the two-level system (TLS) Ref. [69]. A general discussion can be found in Ref. [71].

Figure 3b shows the splitting of the two levels into doublets which are separated by the generalized Rabi frequency, which is a function of the pump-beam intensity. If such a system is now probed by a weak beam in a transmission type of experiment, there are three possibilities to connect the four levels (marked in blue, green, and red). Note that the thickness of the new levels indicates the population of these new eigenstates. Figure 3c shows a transmission spectrum where a pump beam with a Rabi frequency of 36 MHz is detuned by 5 MHz with respect to the bare molecule transition [71]. All the expected features are clearly visible and marked for clarity with the respective colors.

The AC Stark effect can be used to switch the probe beam with about 10 photons per lifetime, displaying a nonlinear effect close to the single-photon level. In Fig. 3d, the switching contrast is shown (red line), calculated as the ratio between the probe transmission with pump on and off. A careful analysis of light collected in the transmission channel reveals another interesting nonlinear phenomenon. If the Fourier transform of the recorded signal is calculated for each pump and probe detuning (as shown in Fig. 3e) one observes, besides the beating of pump and probe (i), another signal which originates from the beating of the probe with light generated with a detuning of $2\delta$ (ii). It stems from a degenerate four-wave mixing process and has about 3% of the signal strength of the probe beam. These experiments impressively demonstrate nonlinear optical processes due to the coupling of extremely weak laser beams to a single molecule.

The long sought after goal of single photon-photon interactions mediated by a molecule is therefore tantalisingly close. The introduction of molecules into nanostructures, discussed in detail in Section 4, can be used to enhance the above described processes and bring us into the regime of creating new quantum devices such as non-demolition photon detectors, deterministic quantum information processing gates, and single-molecule-mediated optical switches.

# 4. Molecule-photon interfaces

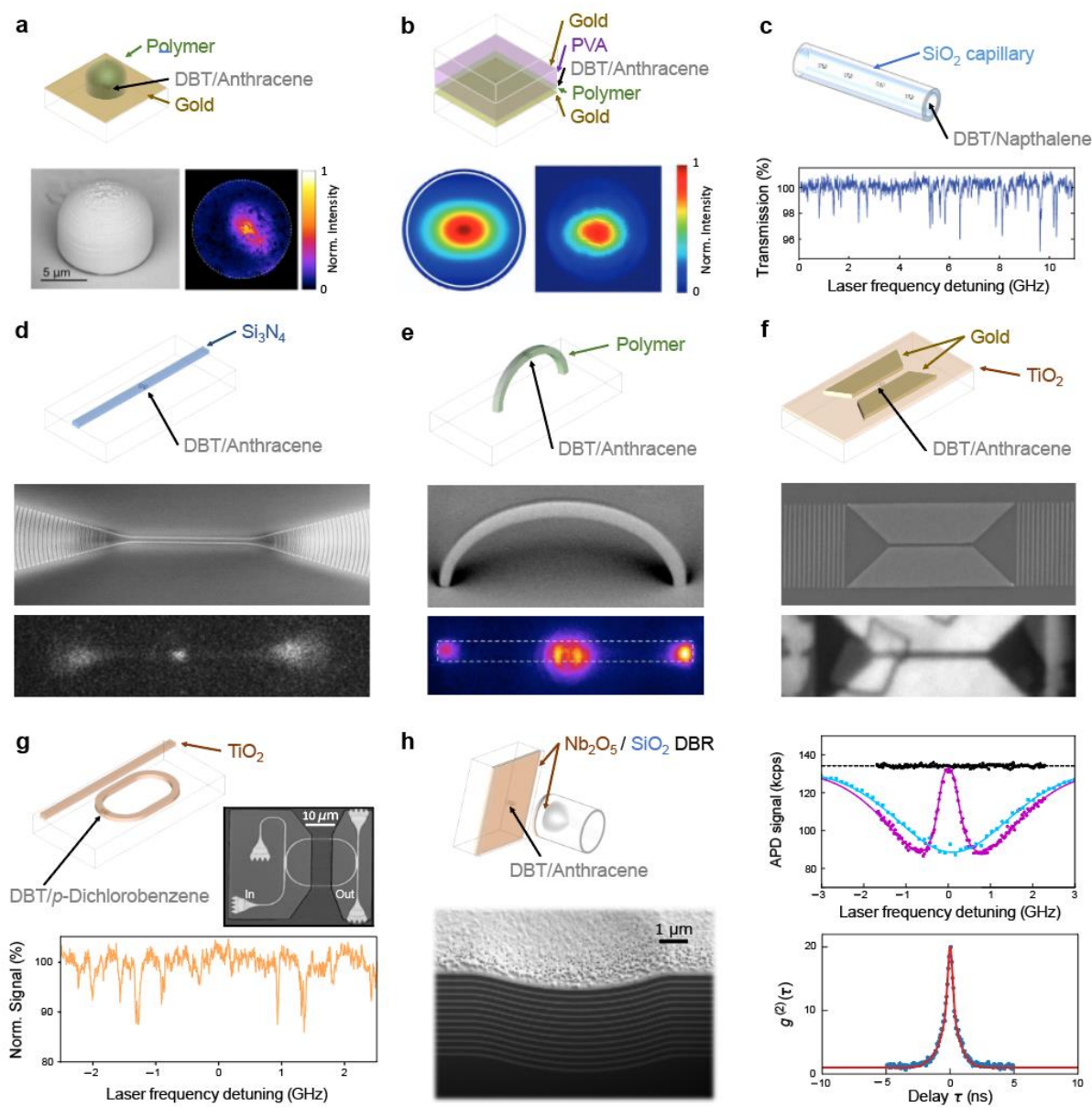

**Fig. 4 | Molecule-photon interfaces. a** Nanofabricated solid immersion lens on gold, used to extract photons from a DBT molecule in an anthracene nanocrystal. Lower left: scanning-electron micrograph (SEM) of the structure. Lower right: Back-focal plane image of DBT fluorescence [49]. **b** A planar dielectric antenna with layers of gold, polymer, DBT-doped anthracene, PVA, and a thin gold director layer. Lower left: simulated angular emission spectrum. Lower right: Back-focal plane image of DBT fluorescence [74]. **c** Silica ($SiO_2$) nano-capillary filled with DBT-doped naphthalene. Lower plot: Transmission through the capillary as the laser frequency is tuned, showing pronounced extinction dips [24]. **d** A silicon nitride ($Si_3N_4$) nanowire waveguide for coupling single molecules to the evanescent field of the guided mode. Center: An SEM image of the fabricated device with gratings. Bottom: fluorescence from a DBT molecule coupled to the waveguide, imaged with an EMCCD camera [63]. **e** A suspended polymer waveguide containing a DBT-doped anthracene nanocrystal. Center: SEM image of the structure. Bottom: fluorescence from an ensemble of DBT molecules, imaged from the underside on an EMCCD camera [49]. **f** Hybrid plasmonic waveguide consisting of gold islands with a varying gap on the surface of a titanium dioxide ($TiO_2$) layer [85]. Center: SEM image of a fabricated device with grating couplers. Bottom: white light microscope image of DBT-doped anthracene crystals on the device. **g** Nanophotonic $TiO_2$ ring resonator used to couple to DBT molecules in para-dichlorobenzene which cover parts of the ring. Center: SEM image of the fabricated device with input/output waveguides. Bottom: Resonant extinction dips as in (c), but with higher visibility due to cavity-enhanced coupling [86]. (**h**) Open-access microcavity formed between flat and rounded distributed Bragg reflectors (DBR) [27]. Bottom left: SEM image of the DBR on the rounded mirror. Top right: Laser reflection from the cavity when it is detuned from both molecule and cavity (black), when it is tuned across the cavity with the molecule detuned, (blue), and when both cavity and molecule are resonant with the laser (purple). Bottom right: Second-order correlation function of the emitted light from the cavity, showing strong photon bunching. Panels **a** and **e** are adapted with permission from [49], Copyright 2020, Wiley-VCH; **g** from Ref. [86] Copyright (2018) American Chemical Society

Although appreciable light-matter interaction can in principle be achieved with strongly focused laser beams in free space [29, 71], tailoring the local electromagnetic field modes available to an emitter is a compact and successful strategy to realize efficient and coherent light-matter interfaces. The relevant parameter describing how much emission from a two-level system (TLS) is collected into a given mode ($\beta$) also determines the coherent extinction of photons in that mode by the TLS. The relative drop in transmission in the absence of dephasing is simply given by $(2\beta - \beta^2)$ [72]. Photonic interfaces are hence crucial for the realization of efficient single-photon sources, as well as enhancing nonlinear interactions at the few-photon level.

Strategies to efficiently extract light from single molecules in free space have evolved from traditional solid-immersion-lens schemes with collection efficiencies from 10% to 20% [73], to nano-fabricated lenses on reflective surfaces (Fig. 4a) addressing a large number of emitters on a single sample [49] and multilayer dielectric and metallic media, tailored for high extraction efficiency. Because molecular crystals have typically lower refractive index with respect to diamond or inorganic semiconductors, even simple, broadband structures allow for efficient extraction of light. Indeed, under ambient conditions, a collection efficiency of 96 % into a 1.65 numerical aperture objective was demonstrated in Ref. [50]. Strong directionality from all emitters on the surface was achieved in a planar antenna (Fig. 4b) [74]. Spatial-mode matching between an incoming laser field and the emission from a single molecule was achieved via coupling to tapered optical fibers [75], fiber facets [76] and sub-wavelength dielectric capillaries [24] (Fig. 4c), with efficiencies around 10% in all cases.

On-chip integration of single molecules represents another important prospect, enabling quantum light sources, processing units and single-photon detectors [77] to be fabricated on the same integrated platform. Individual components have already been demonstrated. These include single molecules evanescent coupling to $Si_3N_4$ and $TiO_2$ waveguides [63, 64, 78], according to the theoretical proposal from Ref. [79]. In Fig. 4d, a fluorescence map from a single DBT molecule on a ridge waveguide is reported, showing light at the output grating couplers, from which the coupling efficiency is estimated to be around 40% [63].

Also, very promising is the direct integration of molecule-doped nanocrystals (see Supplementary Information) in polymeric photoresists, which can be structured around selected emitters, either with electron-beam lithography [80, 81] or by direct laser writing techniques in suspended waveguides [82, 49] (Fig. 4e). Nano-particles containing single molecules have also been deposited by ink-jet printing [83], which allowed controlled deposition on photonic nanostructures. We note in passing that these methods allow for deterministic positioning of molecules and can hence in principle be scaled up to couple several emitters together on the same chip.

Dielectric waveguides enable low-loss propagation of photons, but are limited to wavelength-scale sizes, and therefore also have limited efficiency for coupling to single molecules. To overcome this limit, research was carried out on hybrid plasmonic-dielectric waveguides, which were proposed to improve also single-molecule optical switching [84] beyond the free-space realization [66]. More recently, single molecules were added to hybrid devices, where a coupling efficiency of around 12% was achieved to a propagating mode [85] (Fig. 4f). Such devices can make promising additions to dielectric waveguides, enabling regions of strong light-matter interaction.

Resonant structures can be employed to further enhance the coupling to a single mode. These are crucial for molecules due to the detrimental coupling to vibrational modes, which limits the branching ratio on the coherent ZPL. A cavity can promote decay on the ZPL while suppressing emission through other decay paths, thus enhancing this branching ratio. The coupling of DBT molecules to a nanophotonic titanium dioxide ring resonator was recently demonstrated in Ref.[86]. A coupling efficiency to the cavity mode of ~22% was found, limited by inhomogeneity in the host crystal causing excess scattering of light out of the device. An SEM image together with an extinction measurement is displayed in Fig. 4g.

Open cavities formed of two independently movable mirrors benefit from the ability to selectively align the cavity mode on a molecule of choice [87] and tune the length of the cavity to be on resonance with the molecule. Enhanced interaction of a laser with a single molecule was indeed observed, to the point where the laser could be extinguished in transmission by up to 99% [88, 27] (Fig. 4h). In Ref. [27], the authors achieved a coupling efficiency to the cavity mode of $\beta_{cav} = (97.4 \pm 0.3)$ %, and could estimate a modified branching ratio on the ZPL 95%, starting from typical values 33%.

These results all together demonstrate that an efficient light-matter interaction mediated by cavity structures renders molecules ideal single photon emitters and non-linear elements for integrated quantum photonics. The next steps to use this system as an on-demand photon source is to engineer the cavity to decay into a known optical mode, ideally matched to an optical fiber or waveguide, without compromising any cavity enhancement. Multiple such molecule-cavity systems must then be built and show identical emission. Based on the scalable production of identical molecules, such a system promises to reach world record multi-photon generation efficiencies which will enable photonic quantum technologies that surpass anything developed to date.

At the ensemble level, the integration of organic emitters in microcavities has attracted enormous interest from different communities studying the effects of strong coupling on photochemical reactions on ensemble of molecules [89] and other polaritonic effects. Coupling molecules to the continuum of mode in one-dimensional waveguides or to single-mode cavities appears also an ideal testbed to study many-body coherent effects, such as super- and sub-radiance, as a function of positional disorder, coupling efficiency, inhomogeneous broadening and dephasing [90, 91].

# 5. Single-molecule sensing and quantum sensing

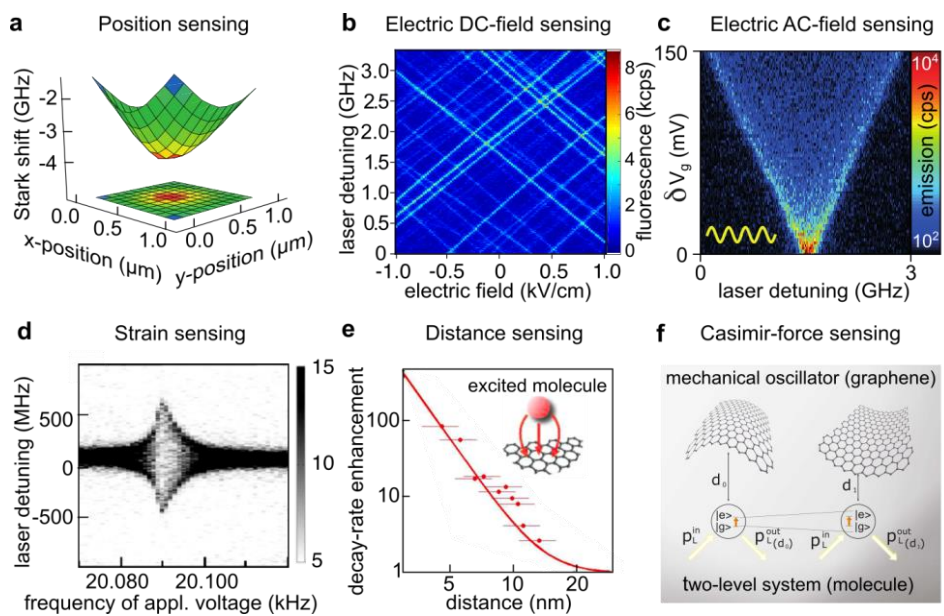

**Figure 5: Using single molecules for quantum sensing. a** Scanning probe microscopy to identify separation and coupling between two terrylene molecules spaced by 12 nm. The scanning tip is terminated with a metallic sphere to apply a non-homogeneous electric field [92]. **b** Linear Stark shift of DBT single molecules in 2,3-dibromonaphthalene; the external electric field was applied nearly parallel to the a-axis of the crystal [60]. **c** Dynamic control of a single DBT molecule's energy through an out-of-plane AC electric field applied through a graphene electrode [61]. **d** Single DBT molecule used as a transducer of local strain field in an anthracene crystal attached to a tuning fork, showing the fork's resonance. At the mechanical resonant frequency, the strain is enhanced and the optical line is broadened [105]. **e** Non-resonant energy transfer between graphene and rhodamine molecules showing decay rate enhancement at small separation, and consequent quenching of the molecular emission under graphene [112]. **f** Theoretical proposal for sensing the mechanical motion of a graphene resonator *via* vacuum potential. The molecule energy levels are shifted depending on the distance to the membrane [116]. The panels are adapted with permissions: **a** from Ref. [92], **b** from Ref. [60], Copyright (2019) **c** from Ref. [61], copyright (2019) American Chemical Society; **d** Ref [105]; **e** from Ref. [112] copyright (2013) American Chemical Society; **f** from Ref. [116] Copyright (2014) by the American Physical Society.

Single fluorescent molecules are well established in microscopy and sensing, where they can act as transducers between light and a local field or deformation. Optical read-out allows for contactless measurements which minimize system perturbations. Another crucial advantage of single molecules for sensing is their nanoscopic size and fixed position inside a solid-state host matrix. Under cryogenic conditions experimental investigation is supported by an in-principle infinite photostability and exquisitely narrow lines, allowing for high spatial resolution sensing. In an electrical equivalent of magnetic resonance imaging, strong electric field gradients were used to resolve two molecules in three dimensions at a resolution of 2 nm [92] (Fig. 5a). There, the massive Stark effect of a nanoscopic probe is harnessed alongside the almost background-free detection under fluorescence excitation of the molecules. Correspondingly, gradients in the optical field can also be sensed by measuring Rabi oscillations [93].

In the early days of cryogenic single-molecule spectroscopy the narrow-band transitions of a molecule were explored to sense external or internal fields which act on the molecule. Investigations were focused on the application of DC-Stark fields [57, 25, 94] and AC-fields [95, 96, 97]. Internal charges in the host matrix can also result in a spectral shift of the molecule's ZPL [98]. Later on, spectral jumps of particular molecules induced by the external laser were observed and the mechanism was suggested as an optical switch [99]. These original observations triggered the idea that a single molecule can act as a transducer, enabling the otherwise impossible optical detection of tiny internal charges, defects and ideally single spins. Such a single-molecule sensor integrates high sensitivity and spatial resolution. The crucial question is how the external influence is mediated towards the molecule, which is usually based on the Stark effect, shifting the ZPL frequency in the electric field.

Towards the goal of single-charge detection, it was shown that the electronegativity of halogen atoms, combined with the herring-bone crystalline structure of a 2,3-dibromonaphthalene host, can induce large electric dipoles on a centrosymmetric molecule such as dibenzoterrylene [60]. The homogeneity and the linearity of the Stark effect in this system can be appreciated from the graph in Fig. 5b, where the excitation spectrum of several molecules is plotted as a function of the applied external electric field. With an electric dipole moment change of about 1 D, this system would allow the optical detection of single electrons at the electron-emitter distance of at least 100 nm and by a full linewidth frequency shift [100, 101]. Similarly, it was proposed in Ref. [102], that a single-charge displacement can be measured by means of single-molecule dynamic triangulation, with positional accuracy better than 6 pm.

External pressure on molecular crystals at low temperatures is another parameter that acts on single chromophores and causes an additional frequency shift of the ZPL. In this way, pressure sensors have been demonstrated with a responsivity of the order of 1 GHz/atm [103] (see also Fig.5c). By exploiting the coupling of local strain to applied AC electric fields, Kol'chenko et al. have used single molecules as nano-microphones, sensing the local vibrations of a solid [104].

Similarly, strain coupling has been used to detect deformations of a micro-tuning fork though the corresponding spectral line shift at low temperatures [105], as shown in Fig. 5d, which demonstrates the interaction with the mechanical oscillator. Coupling light fields to mechanical motion is at the heart of the growing field of optomechanics, which has led to extra-ordinary sensitivity at various length scales and frequency domains (from gravitational wave interferometers to photonics nano-optomechanical sensor devices) and has been readily extended to the quantum domain. Despite their advantages (high mass resolution, high force sensitivity, large scalability), the transduction of motion in nano-mechanical systems remains a challenge. It can be overcome by coupling them with single molecules in the photon-counting regime. A theoretical proposal based on the strong electric-field gradient at the tip of a carbon nanotube moving in the near field of a molecule would enable efficient transduction of the nano-motion as well as a process for optomechanical cooling of the resonator [106]. Recent theoretical advances suggest that the molecule can act as a topological actuator to control the quantum-mechanical spectrum [107].

Compared to electric fields, the sensitivity of transitions between spinless states of single molecules to magnetic fields is limited, and only a weak second-order diamagnetic Zeeman effect can be monitored [108]. This result becomes clear, when it is considered that no unpaired electrons are available in the usual single molecules. The triplet levels of closed-shell molecules are sensitive to first-order Zeeman interactions, and enable spin manipulations by microwave frequencies [109, 110]. There, even coherent oscillations can be generated [111]. The combination of the excellent optical properties of single molecules with electronic and nuclear-spin manipulations in their triplet states would open a rich field for quantum information technologies. The challenge here resides in the difficulty to calculate and measure the singlet-triplet energy spacing, which requires broadly tunable laser systems due to the large range of computational predictions. Such laser sources are nowadays available and such investigation is on-going. The subsequent coherent control of transitions between singlet and triplet states also has yet to be demonstrated, but will no doubt benefit from the interaction of molecules with nanostructures such as those described in Section 4.

Recently, non-resonant energy transfer from single molecules to graphene has paved the way for distance sensing applications [112, 113] and fundamental investigations of near-field interactions between single emitters and 2D materials. The emitter lifetime is strongly affected by the presence of a single graphene layer due to the latter's high mobility, Dirac electronic dispersion, and purely bidimensional nature [114]. The emitter fluorescence yield varies with the distance to graphene and depends solely on the fine-structure constant, the emission wavelength and the distance, which makes fluorescence quenching by graphene a fundamental distance ruler (Fig. 5e) [112]. As a result, one can detect a FRET-like effect to distances beyond the characteristic 10 nm of standard acceptor–donor energy transfer. Such a universal dependence on distance enables precision localization (up to 5 nm) of single DBT molecules in anthracene crystals deposited on top of graphene [113], by measuring the molecule's fluorescence lifetime.

The experiments mentioned so far rely on the specific features of molecules, but most interactions were classical. However, single-molecule sensing can attain single-excitation sensitivity. When a single electron, or a single vibrational quantum is involved, the description becomes quantum mechanical. Entanglement and coherent state transfer between different quantum systems can then be envisaged. In a recent theoretical proposal for instance, the author propose to use moleculs in waveguides to interface superconducting Qubits with optical photons is a hybrid device [115]. Another exciting theoretical proposal that testifies the potential future impact of molecules in quantum sensing is described in Ref. [116]. The authors suggest harnessing vacuum forces (i.e. Casimir-Polder interaction) to dispersively couple a single molecule to the motion of a graphene nano-mechanical resonator (Fig. 5f). As an example, at a distance of 20 nm between the graphene resonator and a DBT molecule, the frequency shift of the DBT molecule is 100 times larger than its natural linewidth. Due to this large coupling, the emitter can be used as a transducer to monitor the motion of the graphene in real-time, or to squeeze the mechanical resonator position.

# Challenges and Outlook

After three decades of hard work on single-molecule studies, their development as efficient and coherent single-photon sources [7, 42, 49,52], the successful demonstration of single-photon-single-molecule interactions [27] including single-photon nonlinearities [71] and their engineering as sensitive probes [92, 105, 60], we are about to embark on an exciting new era. Here, one aims to construct more complex on-chip architectures, where a controlled number of quantum emitters provide different functionalities for a realm of applications in classical and quantum optical technologies. Molecules coupled to photonic nano- and microstructures such as plasmonic nanoantennas, subwavelength waveguides, and microresonators will be particularly promising for exploiting quantum cooperative effects and many-body phenomena that involve linear or nonlinear interactions [117, 90]. Furthermore, quantum optical devices such as nano-sensors based on novel quantum-sensing schemes [105, 116], quantum transducers [115], or multiple sources of non-classical light are within reach.

One of the hurdles to unlock these applications is the control of materials at the single emitter level. While physical manipulations of single molecules were reported as early as the late 1990s, nanometer control in coupling single molecules to photonic structures remains a topic of current research [49, 83]. Future work will have to develop chemical and physical methods to tailor organic and hybrid materials with single-molecule precision. These efforts will be crucial for a number of exciting quantum technological applications, where the efficiency of light-matter interaction phenomena plays an important role [118].

Another central matter of concern, as for any solid-state optical emitter, is host matrix interactions. Indeed, it is the influence of phonons that limit many molecular properties such as the quantum yield, decoherence, absorption cross section, and spectral stability. To tame these unwanted interactions one is often obliged to perform cryogenic measurements. Despite their key role, however, phonon couplings have often been avoided in solid-state quantum optics rather than scrutinized. Future studies will aim at investigating the "opto-mechanical" degrees of freedom both within a single organic molecule (vibrational transitions) and with its environment (optical and acoustic phonons) [119, 21, 22]. In addition to minimizing unwanted couplings to improve the degree of coherence at elevated temperatures, there is promise for exploiting the rich phononic landscape of molecules as a resource.

Many of the protocols developed for quantum information processing rely on so-called $\lambda$ configurations, where two long-lived ground states are connected via an excited state. However, common organic dye molecules possess one long-lived ground state because their electron orbitals are composed of paired electrons, so that the ground state is typically a singlet state with no net electronic spin. Here further investigations, of radicals for example, could open new doors to molecular systems with novel optical properties, including coupling to spins in doublet or triplet ground states. Furthermore, such chemical developments would enable magnetic field sensing at the nanoscale.

Organic molecules are often associated with insufficient photostability since their most commonly known applications are in the pigment industry and fluorescence microscopy, where photobleaching is omnipresent. In the context discussed here, this impression is unfounded in many ways because optical excitation of organic molecules in the 1-2 eV energy range is far from the direct photodamage threshold, and photostability is only endangered if chemical reactions are possible. Chemical modifications are strongly inhibited at cryogenic temperatures. Moreover, such long-term photostability can be attained also at room temperature, for samples encapsulated under a controlled atmosphere. Indeed, while the use of organics for large display screens seemed unthinkable at the end of the 1990s, controlled fabrication and packaging have allowed the industry to overcome the photostability issue. Similar engineering can be envisaged for single-molecule-based quantum devices.

Future high-performance optical quantum technologies will most likely require hybrid platforms to benefit from the best features (optical, electrical, magnetic, mechanical, etc.) of various materials and qubits. Considering their favorable properties such as high quantum yield, ease of fabrication and handling, small size and availability at different wavelengths, as well as brightness and high degree of coherence, molecules are ideally suited for such endeavors [115]. What is, indeed, particularly intriguing is that a molecule is intrinsically a hybrid system in which electronic and nuclear degrees of freedom provide a broad range of transition frequencies in the optical, infrared, and microwave domain. The exciting path towards harnessing molecules in quantum technologies will most certainly also involve new discoveries that we might be overlooking today.

## Acknowledgments

This project has received funding from the EraNET Cofund Initiatives QuantERA within the European Union's Horizon 2020 research and innovation program grant agreement No. 731473 (project ORQUID). ASC acknowledges a University Research Fellowship from the Royal Society (UF160475). W.F. and I.G. acknowledges fundings from the Deutsche Forschungs gemeinschaft (DFG) - Projektnummer 332724366 and GE2737/5-1, respectively. Amin Moradi is thanked for discussions and NWO (The Dutch Research Council) for funding of his PhD grant on sensing of single charges. C.T. wishes to thank Alois Renn for always useful discussions.

## Conflict of Interest

The authors declare no conflict of interest.

# Supplementary Information

# Single organic molecules for photonic quantum technologies

C. Toninelli[1], I. Gerhardt[2], A.S. Clark[3], A. Reserbat-Plantey[4], S. Götzinger[5,6], Z. Ristanovic[7], M. Colautti[1], P. Lombardi[1], K.D. Major[3], I. Deperasińska[8], W.H. Pernice[9], F.H.L. Koppens[4], B. Kozankiewicz[8], A. Gourdon[10], V. Sandoghdar[5,6], and M. Orrit[7]

1 CNR-INO and LENS, Via Nello Carrara 1, 50019 Sesto Fiorentino (FI), Italy
2 Institute for Quantum Science and Technology (IQST) and 3rd Institute of Physics, D-70569 Stuttgart, Germany
3 Centre for Cold Matter, Blackett Laboratory, Imperial College London, Prince Consort Road, SW7 2AZ, London, United Kingdom
4 ICFO − Institut de Ciencies Fotoniques, The Barcelona Institute of Science and Technology, 08860 Castelldefels (Barcelona), Spain
5 Max Planck Institute for the Science of Light, 91058 Erlangen, Germany;
6 Friedrich-Alexander University of Erlangen-Nürnberg, Germany;
7 Huygens-Kamerlingh Onnes Laboratory, LION, Postbus 9504, 2300 RA Leiden, The Netherlands
8 Institute of Physics, Polish Academy of Sciences, Al. Lotnikow 32/46, 02-668 Warsaw, Poland
9 Physikalisches Institut, Westfälische Wilhelms, Universität Münster, Heisenbergstrasse 11, 48149 Münster, Germany
10 CEMES-CNRS 29 Rue J. Marvig, 31055 Toulouse, France

## Sample types and preparation

There exist several techniques to produce molecular samples, all of which do not require high-end machines such as molecular beam epitaxy or ionic/electronic beams. The fabrication processes behind all the experiments discussed so far are cost effective and fast. The choice of a specific preparation method depends on the requirements of the single-molecule experiment. The homogeneous linewidths, inhomogeneous broadening, spectral diffusion and response to an external field are some of the photophysical aspects that need to be considered. Other properties that need to be defined, playing important role especially for the integration of molecules in photonic devices, are the concentration, orientation, and positioning of single emitters.

**Host materials.** To provide desirable photophysical properties, fluorescent emitters (guest molecules) are embedded in a matrix consisting of suitable host molecules. For this reason, the outcome of many experiments may critically depend on the purity of the host material. Zone-refined compounds of the highest purity are used for this purpose [schwoerer2007]. In some cases, when zone refining of the host is not possible, other methods of crystal purification should be considered. For example, column chromatography, recrystallization, or (gradient) sublimation may be helpful to remove some impurities from the host material [karl1980, schwoerer2007].

**Single crystals.** The highest-quality single crystals for photonic quantum application are obtained by sublimation of zone-refined material under an inert gas atmosphere, with typical pressure around 200 mbar. The sublimation temperature is stabilized 10-20 degrees below the materials' melting point and single crystals are collected at a plate kept at slightly lower temperature. An extra heating of a small crucible filled with the fluorescent single molecules, as a part of the sublimation device, allows fine control of the dopant concentration, even up to very high concentrations [major2015, bialkowska2017]. Preferred are crystals which grow in form of thin plates of a few micrometers in thickness, like those of naphthalene, anthracene or *p*-terphenyl. Well-defined substitution sites occupied by the dopants in these crystals allow for the precise orientation of the molecules with respect to the experimental set-up, as these are often placed in an optical confocal microscope configuration [moradi2019]. When a co-sublimation method is not suitable, the Bridgman method can be used to grow high-quality single crystals containing guest molecules in low concentrations [orrit1990, jelezko1996]. An important advantage of using single-crystalline samples for photonic quantum technologies is their high resistance to photobleaching and photoblinking, allowing for prolonged investigation of the same single dye molecule (see Section 2).

**Self-assembled nanocrystals.** Another technique for sample preparation, which has recently found multiple applications is reprecipitation of nanocrystals from solution. Nanocrystals of anthracene doped with DBT have been obtained by the method described in Ref. [pazzagli2018]. A dilute solution of DBT and anthracene is prepared in acetone (or another water-miscible solvent) which is subsequently injected into sonicated water. As the resulting microdroplets of acetone slowly dissolve in water, DBT and anthracene reprecipitate from the solution in the form of nanocrystals, typically hundreds of nanometers in size. Nanocrystals of anthracene doped with DBT can be deterministically integrated, for example, in polymer-based 2-D and 3-D photonic structures [ciancico2019, colautti2020]. With this method, the final concentration of the dopant is accurately controlled and the background fluorescence is minimized. The addition of methyl-methacrylate monomers and a polymerizing agent to this process results in nanocrystals that are covered with a poly-methyl methacrylate protective shell which prevents sublimation [schofield2020].

**Polycrystals.** This method is the simplest to produce single-molecule samples with a large range of concentrations and without the need for special preparation procedures. Shpol'skiĭ matrices are common hosts to observe narrow zero-phonon lines (ZPLs) in many systems [shpolskii1960, brunel1999]. Guest molecules are dissolved and diluted in a host, which is usually a liquid. Low-molecular-weight normal alkanes are employed as host molecules, because they crystallize when frozen and their absorption occurs at higher energy than that of all $\pi$-$\pi$ electronic transitions in PAHs. They hence interact weakly with the laser and the chromophores. The length of the alkanes is chosen to approximately match at least one of the dimensions of the chromophore, and are usually in the size range between n-pentane and n-dodecane. The sample is then cooled spontaneously or at a desirable cooling rate. Low spectral diffusion and high photostability of single emitters can be achieved for instance for DBATT in *n*-tetradecane [rezai2018]. Suitable smaller PAHs can also be used to generate polycrystalline samples from the melt. This method was used for instance in Ref. [faez2014a], providing narrow and stable emission from DBT in naphthalene.

For certain applications, such as sub-shot-noise imaging, narrow linewidths are not required and single-molecule emitters can be operated at room temperature. In this case, molecular polycrystalline samples can be produced by spin-casting a droplet of solution, supersaturated with the host molecule and with a controlled concentration of the dopant chromophore, on a rapidly spinning plate. Examples of experiments employing spin-coated samples include those described in Refs. [toninelli2010a, pfab2004]. The basic advantage of this method is that the produced crystals are few-tens-of-nm thick, which reduces the background and allows near-field investigations.

# The choice of a chromophore: state of the art and outlook

The number of molecules so far studied for single-molecule photophysics and quantum applications in particular is somehow small but certainly could increase in the next decade. To date, most experiments have been limited to a few polyaromatic hydrocarbons (PAH): pentacene, perylene, terrylene, dibenzoterrylene (DBT), dibenzanthanthrene (DBATT), as shown in Fig.1f and perylene bisimide. The constraints on the design of new molecular emitters appropriate for quantum technologies are rather strong:

a) suitable electronic properties, as discussed in Section 1, in particular strong transition probability into the ZPL and small Stokes shift, which implies rigid molecules. Indeed, apart from one exception [kiraz2005], all single molecules with strong Fourier-limited ZPLs known so far are extremely rigid and flat PAHs.

b) absorption wavelengths in the red to infrared to fit usual laser wavelengths and to reduce fluorescence from impurities and optics. In the case of PAH, these small HOMO-LUMO gaps are obtained by increasing the number of fused benzene rings while limiting the number of Clar sextets. [Sola2013] Further extension of the delocalized $\pi$-system, such as in oligorylenes [Avlasevich2006], [Langhals2015], or oligoacenes can lead to absorbances in the infrared around 1 micron. This wavelength range is interesting for coupling with silicon photonics and for quantum communications in general. However, molecules absorbing at such wavelengths present low fluorescence efficiencies as a consequence of the energy gap law [Siebrand in Birks]. For smaller HOMO-LUMO gaps, fewer vibrational quanta are involved in internal conversion, and the Franck-Condon factors become exponentially larger. Creation of a few C-H (or N-H, O-H) stretch quanta competes efficiently with spontaneous emission. A possible remedy is to deuterate the molecules, so as to lower the vibrational quantum of C-D (or N-D, O-D) bonds. Another solution to this difficulty could be to substitute peripheral hydrogen atoms by other rigid substituents like halides [sakamoto2006].
Another challenging point is that reducing the HOMO-LUMO gaps increases the reactivity of molecules towards photo-oxidation or dimerization. For instance, acenes longer than pentacene only exist as dimers in solution.

c) negligible spectral diffusion implying that these PAH cannot be substituted by non-rigid groups with internal degrees of freedom (see remark in (a) above). This constraint is a severe one since the addition of bulky and flexible substituents is the way to improve the solubility of rigid planar molecules by increasing the intermolecular distances, ie reducing the $\pi-\pi$ stacking, and also to protect them from further reactions. This lack of solubility of unsubstituted PAH implies that their synthesis, and often that of their precursors, is carried out in harsh conditions, that the purification steps are problematic, often limited to washing and sublimation when possible, and that standard spectroscopic characterizations like NMR in solution are limited.

A possible alternative new way of studying large, insoluble, electronically delocalized and reactive unsubstituted PAH is first to prepare soluble and chemically stable non-planar precursors that can be synthesized, fully purified and characterized by standard in-solution organic chemistry techniques. These compounds can then be diluted in host materials. In the final step, the volatile protecting groups can be removed by UV irradiation, yielding the final PAH as shown below in the case of long acenes [Yamada 2005, watanabe 2012, Jancarik 2019]. In the example of Fig.S1, the precursors are non-planar carbonylated compounds, comprising small units of large HOMO-LUMO gaps like naphthalene or anthracene sub units. The presence of the bridging carbonyl group prevents strong intermolecular π−π interaction and breaks the electronic coupling between sub-units. After dilution in a matrix, photodeprotection removes volatile carbon monoxide leaving isolated insoluble PAH.

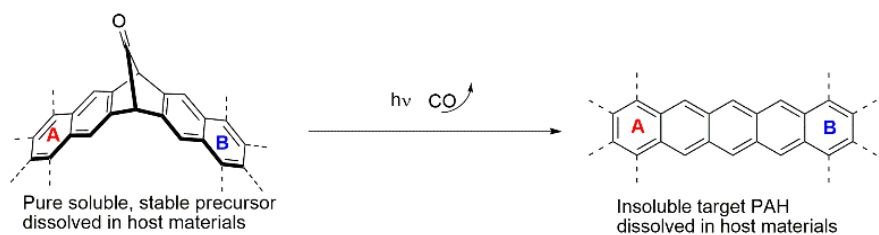

**Fig.SI. 1** In-situ preparation of PAH diluted in a solid matrix by photochemical decarbonylation of a precursor.